\documentclass[conference]{IEEEtran}
\IEEEoverridecommandlockouts
\usepackage{cite}
\usepackage{amsmath,amssymb,amsfonts}
\usepackage{algorithmic}
\usepackage{graphicx}
\usepackage{textcomp}
\usepackage{xcolor}

\usepackage{algorithm}
\usepackage{subcaption}
\usepackage{hyperref}
\usepackage[compatibility=false]{caption}

\def\BibTeX{{\rm B\kern-.05em{\sc i\kern-.025em b}\kern-.08em
    T\kern-.1667em\lower.7ex\hbox{E}\kern-.125emX}}
\begin{document}

\title{MusicAIR: A Multimodal AI Music Generation Framework Powered by an Algorithm-Driven Core}

\renewcommand{\footnoterule}{%
  \kern -3pt
  \hrule width 0.2\textwidth
  \kern 2.6pt
}

\author{
\IEEEauthorblockN{Callie C. Liao}
 \IEEEauthorblockA{
\textit{Stanford University}\\
Stanford, USA \\
ccliao@stanford.edu}
 \and
\IEEEauthorblockN{Duoduo Liao}
 \IEEEauthorblockA{
 \textit{George Mason University}\\
Fairfax, USA \\
dliao2@gmu.edu}
 \and
\IEEEauthorblockN{Ellie L. Zhang}
 \IEEEauthorblockA{ \textit{IntelliSky}\\
McLean, USA \\
elzhang@intellisky.org}
}

\maketitle

\begin{abstract}
Recent advances in generative AI have made music generation a prominent research focus. However, many neural-based models rely on large datasets, raising concerns about copyright infringement and high-performance costs. In contrast, we propose MusicAIR, an innovative multimodal AI music generation framework powered by a novel algorithm-driven symbolic music core, effectively mitigating copyright infringement risks. The music core algorithms connect critical lyrical and rhythmic information to automatically derive musical features, creating a complete, coherent melodic score solely from the lyrics. The MusicAIR framework facilitates music generation from lyrics, text, and images. The generated score adheres to established principles of music theory, lyrical structure, and rhythmic conventions. We developed Generate AI Music (GenAIM\footnote{GenAIM: \url{https://genaim.intellisky.ai/}}), a web tool using MusicAIR for lyric-to-song, text-to-music, and image-to-music generation. In our experiments, we evaluated AI-generated music scores produced by the system using both standard music metrics and innovative analysis that compares these compositions with original works. The system achieves an average key confidence of 85\%, outperforming human composers at 79\%, and aligns closely with established music theory standards, demonstrating its ability to generate diverse, human-like compositions. As a co-pilot tool, GenAIM can serve as a reliable music composition assistant and a possible educational composition tutor while simultaneously lowering the entry barrier for all aspiring musicians, which is innovative and significantly contributes to AI for music generation.

\end{abstract}


\section{Introduction}

Generative AI has experienced rapid escalation and integration into daily lives, particularly with the frequent use of conversational chatbots such as ChatGPT \cite{achiam2023gpt} powered by Large-Language Models (LLMs). However, AI music generation, especially multimodal inputs from images, lags behind AI art and writing due to its complex structure and required musical expertise. Current methods rely on deep learning \cite{agostinelli2023musiclm}\cite{Chen2023MusicLDMEN}\cite{xmusic2025}\cite{LeCun2015Deep}, using large datasets for music generation in formats such as song scores, audio, plain text, or other types of data, but face challenges including data collection, copyright risks, high computing costs, and labor-intensive data preparation \cite{NEURIPS2023_94b472a1}\cite{Bao2018NeuralMC}\cite{dong2018convolutional}. \cite{Ji2020ACS} also describes a comprehensive survey regarding deep learning methods for music creation. These approaches pose several challenges: 1) significant amounts of data need to be collected to produce more accurate results; 2) potential copyright infringement issues for generated music can arise while training models; 3) high-cost computing resources may be required for training, limiting researchers' capabilities in producing reliable models for music generation; and 4) most models involve significant data collection and processing, which would frequently necessitate heavy manual labor.   

\begin{figure}
 \centerline{
 \includegraphics[width=0.9\columnwidth]{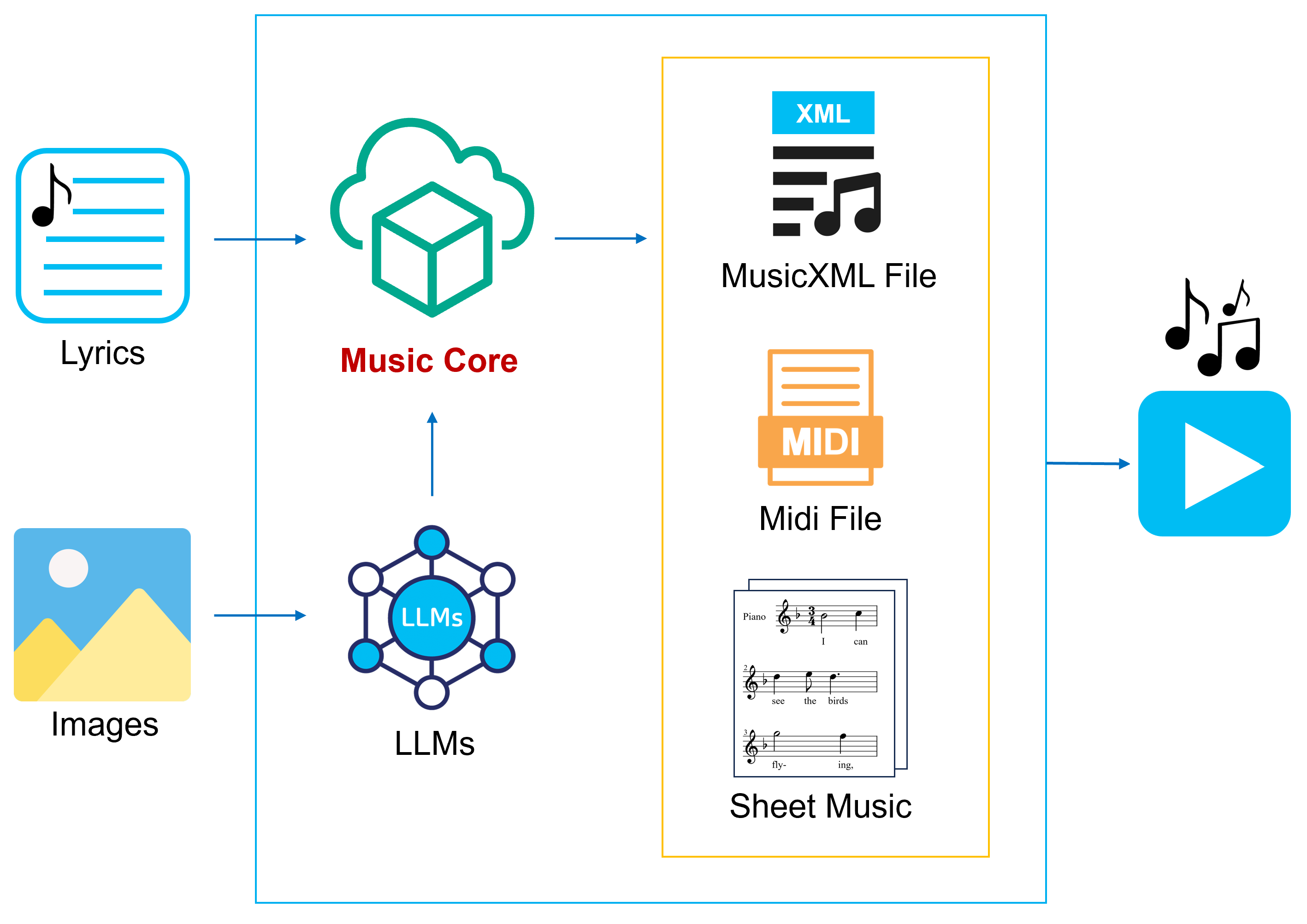}}
 \caption{The framework of MusicAIR.}
 \label{fig:MusicAIR_sys}
\end{figure}

\begin{figure*} 
     \centering
     \begin{subfigure}{\textwidth}
         \centering
         \includegraphics[width=1\linewidth]{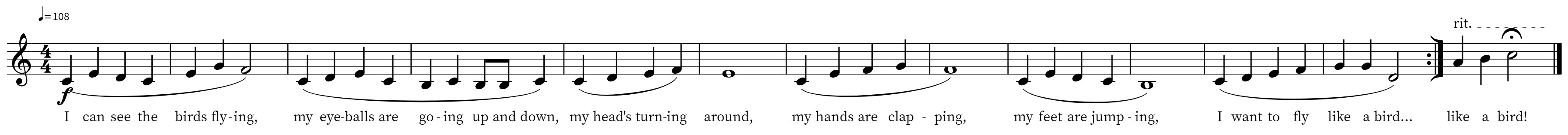}
         \caption{Original composition \cite{birdsareflying}.}
         \label{fig:EllieLZ_score_strip}
     \end{subfigure}    
    \hfill
     \begin{subfigure}{\textwidth}
         \centering
         \includegraphics[width=1\linewidth]{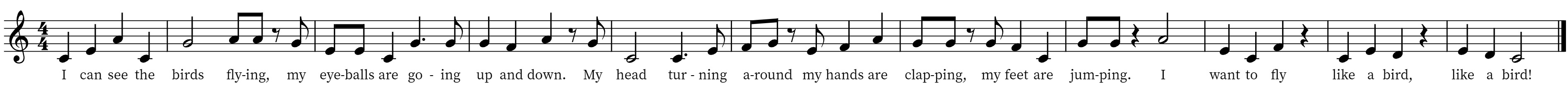}
         \caption{AI-generated song.}
         \label{fig:AI_score_strip}
     \end{subfigure}
        \caption{AI-generated song vs. human-written song using the same lyrics.}
        \label{fig:Two compositions}
\end{figure*}

There has been progress made in pure music generation \cite{NEURIPS2023_94b472a1}\cite{huang2023noise2music}\cite{agostinelli2023musiclm}, but lyrics-based music generation has been more difficult. Recently, there has been an increase in text-guided music generation research such as text-to-melody, text-to-music, and text-to-audio generation, where text is not included as lyrics, contributing to pure music generation \cite{NEURIPS2023_94b472a1} \cite{agostinelli2023musiclm} \cite{Chen2023MusicLDMEN}. However, most of the methods described above are neural-based, and even though many of them incorporate some music theory knowledge and conventions in an effort to create more reasonable models, it could raise the possibility that their generated music diverges from music theory guidelines and proper lyric-note alignment. Additionally, it is possible that utilizing existing songs as data introduces potential copyright infringement issues in the future. 

In developing a musically robust song generation method, knowledge of music theory, literature, and linguistics is required to emulate the intuitive thinking process of lyricists, composers, and singer-songwriters, while not necessarily relying on existing music to train and develop such algorithms. Therefore, we propose Music AI fRamework (MusicAIR), an innovative multimodal framework powered by a novel purely algorithm-driven symbolic music core. Our core method only utilizes lyrics and novel algorithms, addressing the challenges listed above. We aim to leverage the correlations between lyrical and musical components, as inspired by \cite{Liao2022Multimodal}\cite{CLiao2024} where through quantitative analytics, keywords are discovered to have a tendency to land on strong beats to create songs with rhythmic structures suitable for both the lyrics and the melodies. Our approach also attempts to discover latent rhythmic, syllabic, and stress patterns to help determine the rhythm and pitches. The lyrics and the melodies would be properly aligned, with lyrical stress matching the emphasized beats (i.e., strong beats) in the music. 

For this pilot study, we created Generate AI Music (GenAIM), a lyric-to-song, image-to-music, and lyric-to-music generation web tool based on our MusicAIR framework. Figure \ref{fig:lyric_image_gen} presents two examples of GenAIM-generated music. Our generation method includes features such as customizable key signatures and instruments for playback (Figure \ref{fig:framework_ai_music_user}), as well as sheet music for display (Figure \ref{fig:genAIM_music history}). In our analysis, we used music-theory-based metrics, as evaluating the degree of alignment to music theory guidelines for the generated melodies is our core focus.

In this paper, we encompass the relationship between lyrics and melody, music theory guidelines, and composers' intuition to develop a non-neural approach towards AI music generation from lyrics and generate human-sounding music, which differentiates our approach from existing methods in this field.

Our main contributions are listed as follows: 
\begin{itemize} 
    \item Our framework, MusicAIR, processes various input types, including lyrics and images, to generate pleasant-sounding melodies utilizing only input lyrics or text derived from images. This innovative approach, particularly the novel non-neural lyric-to-music algorithms that does not rely on training data, drastically differentiates from existing music generation models while ensuring high lyric-music compatibilities. It ensures the avoidance of copyright infringement while remaining cost-effective.
    \item Our key methods in the music core for song generation is achieving lyrical alignment with the rhythmic structure through the innovative use of keywords and strong beats for proper beat alignment, alongside an emphasis on music theory conventions.
    \item The image-to-music methods proposed in our MusicAIR framework employ Large Language Model (LLM) technologies to only generate lyrics, followed by the use of our purely algorithmic approach to automatically compose classical music or songs.
    \item As a co-pilot web tool, GenAIM inspires composers and supports individuals of any background by utilizing visual imagery to transform thoughts to music, all while preserving human creativity as a reliable music composition assistant. Moreover, with the addition of symbolic music generation and the flexibility of melody generation, GenAIM can be highly useful in education by serving as a composition tutor to students of any level given that sheet music and audio replay are provided. Additionally, it offers a range of benefits, including entertainment, relaxation, and support for mental well-being.  
    \item Our evaluations are primarily based on music theory rules instead of human listening feedback in other research concerning music generation, as music theory standards are objective and thus more fitting for the research in this paper than listening evaluations, which the latter would come from varied music backgrounds and mixed perceptions, making it difficult to consolidate. We innovatively apply evaluation metrics such as interval-based melodic smoothness and step ratios to evaluate AI-generated songs based on comparisons with the original human-composed songs, introducing more robust theory-based evaluations to the field of AI music. 
\end{itemize} 

\section{Related Work}\label{sec:related work}

\subsection{Lyric-to-Music Generation}

There has been progress made, particularly in lyric-to-music generation with or without neural networks. Some have been developed into more efficient music generation methods, although there are limitations. 

\subsubsection{Lyric-to-Music Generation with Neural Networks}
Recent research has investigated the generation of music from lyrics using neural networks techniques, especially deep learning. \cite{Bao2018NeuralMC} develops the melody composition model based on the sequence-to-sequence framework to compose melody from lyrics. \cite{Sheng2020SongMASS} proposes SongMASS that expands upon the previous MASS \cite{song2019mass} by switching to a song-masked pre-training strategy and a separate lyric-to-lyric and melody-to-melody encoder-decoder to ensure smoother lyric-to-melody and melody-to-lyric generation. \cite{Ju2022TeleMelody} proposes a two-melody self-supervised generation system that is neural-based and utilizes paired lyric-rhythm data. Some researchers have incorporated music theory and relationships between the lyrics, rhythm, and melody as well \cite{Zhang2022ReLyMe}  \cite{zhang2023modeling}. Although these approaches have achieved relative reliability in their results and improvements, the models rely on significant amounts of prior training data, potentially increasing the computing power required to process large datasets. It also raises the question of improving neural-based models in modern-day AI, which is also related to scaling laws \cite{kaplan2020scaling}: is collecting more data the only method forward for technological advancements? 

\subsubsection{Lyric-to-Music Generation without Neural Networks} 
Few non-deep learning methods for lyric-to-music generation have been investigated. \cite{Long2013TMusic} proposes a purely algorithmic composition algorithm that uses lyric-note correlations captured by frequent pattern mining on the song database to generate Probabilistic Automaton (PA) \cite{RABIN1963230} for melody composition. This method relies on data, neglecting pitch-lyric matching and alignment, and only considers tonal languages, raising concerns about stress and syllable grouping. \cite{Scirea2015SMUGSM} introduces Markov chains \cite{Ames2017TheMP} for lyrics-to-melody generation, though challenges exist in maintaining high-level consistency due to local statistical similarities. \cite{Fukayama2010AutomaticSC} introduces an algorithm that considers compositions in a method inclined towards using probability to achieve music generation, but there is not a strong emphasis on lyrical alignment and specific consideration on lyrical and language aspects such as stress.

We aim to directly generate a song from lyrics by using pure algorithms geared towards a composer's intuition as well as non-neural-based approaches. Furthermore, we creatively leverage the relationship between keywords, stressed syllables, and stressed beats, as well as other statistical, rhythmic, and lyrical features found in \cite{Liao2022Multimodal} \cite{Liao2023AutomaticTS} \cite{CLiao2024}, all of which are derived from patterns in music theory. The derived patterns are largely based on the intuition of composers and lyricists, mirroring our approach for the music core. 

\setlength{\parskip}{0em}    

\subsection{Image-to-Music Generation}
Currently, image-to-music generation remain relatively unexplored. Research in this area involves deep learning methods that extract essential visual attributes, which are subsequently translated into melodies, harmonies, and rhythmic patterns \cite{img2music1}. Emotion is heavily considered in various ways. Some utilize the Valence-Arousal (VA) emotional space to detect the emotional tone of an image, while other researchers perform image analysis while assuming that components are musically related \cite{wang2024emotion} \cite{hisariya2024bridgingpaintingsmusic} \cite{sergio2015genmusicimage}. In general, all the works emphasize the potential emotions that could be extracted from the images. Recent research methodologies \cite{CVPR2024_img2music} \cite{MusicGenDL_Review2025}\cite{CVPR2024_video2music}  focus on generating audio music from multiple modalities, including images, and text. MelFusion \cite{CVPR2024_img2music} synthesizes music from image and language cues using diffusion models. However, our image-to-music methodology distinguishes from these by utilizing LLM technologies to generate lyrics from an input image before employing the music core, our purely algorithmic lyric-to-music approach, to generate the corresponding music, reducing the need to specifically analyze emotions from the image.

\section{The Methodology}\label{sec:page_size}

The system architectures consist of the MusicAIR framework and music core, each of which is described in detail in the following sections.

\subsection{The Framework of MusicAIR}

As shown in Figure \ref{fig:MusicAIR_sys}, the MusicAIR framework encompasses a range of inputs, including textual lyrics and images. Lyrics are directly fed into the music core, which creates music from the text input, generating both sheet music and audio files. Images are first converted into lyrics using LLMs \cite{IMAiGen2025} before further processing. The generated lyrics are then passed to the music core, where elements such as pitches, rhythms, and phrases are determined based on the text. Once the music or song is generated, the framework outputs the results in MIDI or sheet music format. 
The specific algorithms used in the music core are expanded upon below.

\subsection{The Music Core }
\begin{figure}[h!]
 \centerline{
 \includegraphics[width=0.95\columnwidth]{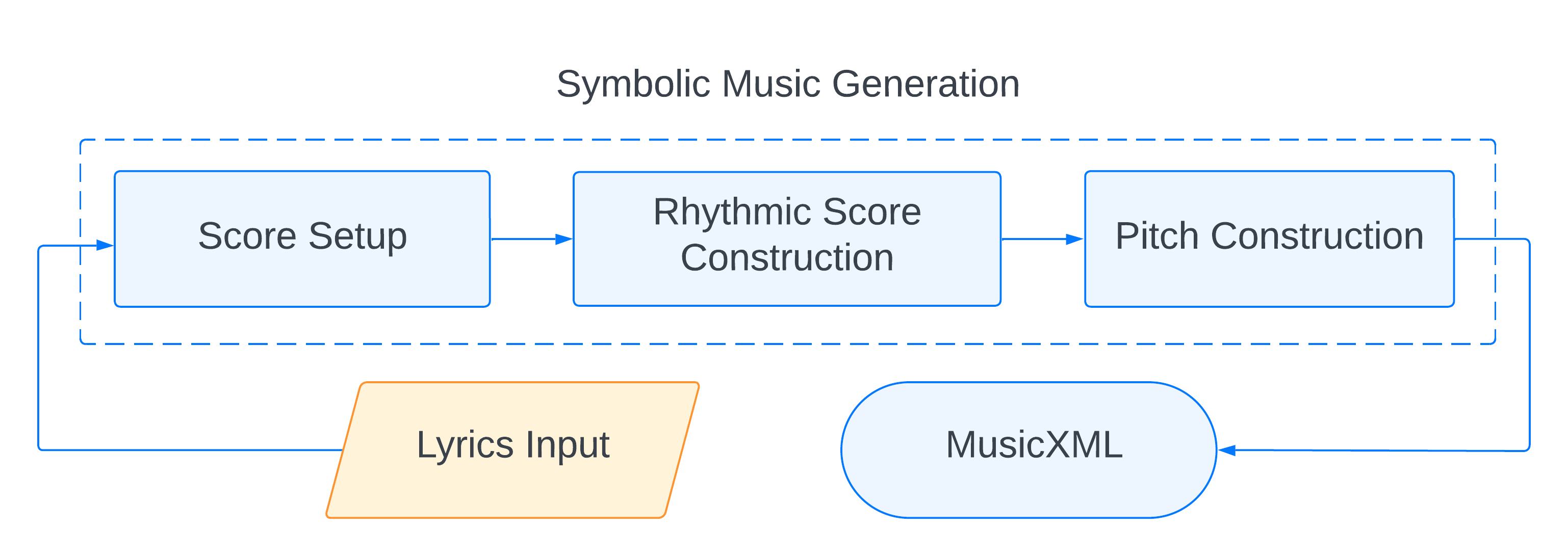}}
 \caption{The flowchart of the music core.}
 \label{fig:music_core}
\end{figure}

The innovative AI music generation core, comprises four key modules: score setup, rhythmic score construction, pitch construction, and score conversion as shown in Figure \ref{fig:music_core}. Each of these modules is explained in the following sections.

\subsubsection{Score Setup}

To set up the score, lyrical information is first extracted from the lyrics, with key information including syllables and keywords prioritized. Phrases are identified primarily through punctuation and sentence endings. Keyword and syllabic information are utilized to determine the time signature for the new score through the method described in \cite{Liao2023AutomaticTS}, which uses machine learning models, or through pure algorithms that extract latent stress information directly from lyrical patterns. After the time signature is determined, the number of accents per measure is retrieved. Along with the number of keywords, this becomes one of the factors for determining the total number of measures. For the key signature, the sentiment feature is used to roughly determine whether the key signature would be major or minor. Sentiment analysis is conducted to extract the positivity, negativity, or neutrality of the lyrical content. Generally, positive sentiment aligns with the major key and negative sentiment aligns with the minor key. However, users are able to select their preferred key signature. The score setup algorithm is described in Algorithm 1.

\noindent\rule{8.cm}{0.4pt}

\textbf{Algorithm 1:} Score Setup

\noindent\rule{8.cm}{0.4pt}

\textbf{Input:} lyrical text of one song $T$.

\textbf{Output:} time signature $ts$, key signature $ks$,

 \qquad \qquad   number of measures $Nm$, 
 
 \qquad  \qquad  a phrase list $Lp$, a word dictionary $Dw$.

1: Initialize $Dw$

2: Create a nested list $Lp$ from $T$

3: \textbf{for} each phrase $p$ in $Lp$ \textbf{do}

4:    \qquad \textbf{for} each word $w$ in $p$ \textbf{do}
            
5:        \qquad  \qquad $flag$ = DetermineKeyword($w$) 

6:  \qquad  \qquad $sylls$ = GetSyllables($w$)
        
7:        \qquad  \qquad $Dw$[$w$] = $tuple(flag, sylls)$ 
            
8:   \qquad  \textbf{end for}
    
9:  \textbf{end for} 
  
10: sent = GetSentiment($T$)  

11: $ts$, $ks$ = DetermineSignatures($Lp$, sent)  

12: $Nm$ = DetermineMeasureNum($Dw$, $ts$) 

13: \textbf{return} $ts$, $ks$, $Nm$, $Lp$, $Dw$

\noindent\rule{8.cm}{0.4pt}


\subsubsection{Rhythmic Score Construction}\label{subsec:rhythm is the best way to go}
The time signature is critical for constructing the rhythmic structure of the score, establishing the fundamental structure of the song, and ensuring each phrase is distributed into the appropriate number of measures. Keyword insertion is then performed by inserting keywords into the stressed beats within each measure following the order specified by the lyrics. The connection between keywords and stressed beats, as discovered in \cite{Liao2022Multimodal} \cite{CLiao2024}, is employed in this proposed lyric-to-music generation framework. Algorithm 2 describes the method.

\noindent\rule{8.cm}{0.4pt}

\textbf{Algorithm 2:} Rhythmic Score Construction

\noindent\rule{8.cm}{0.4pt}

\textbf{Input:} time signature $ts$, number of measures $Nm$, 

\qquad  \qquad a phrase list $Lp$, a word dictionary $Dw$.

\textbf{Output:} A rhythmic score list $Lr$.

1: Initialize $Lr$ as a list 

2: \textbf{for} each measure \textbf{do}
    
3:  \qquad $R$ = BuildRhythm($Dw$, $Lp$)

4: \qquad Insert $R$ into $Lr$

5: \textbf{end for}

6: \textbf{return} $Lr$

\noindent\rule{8.cm}{0.4pt}

\subsubsection{Pitch Construction}\label{subsec:pitch your business now}
When all the lyrics have been placed in their corresponding locations, the rhythmic structure is established for the song, and the pitch is considered. The parameters for the pitch construction are set up first. The key signature, pitch range, and phrase lengths are considered. The key signature not only establishes the mood and tone of the song, but also the appropriate notes for the major/Ionian or minor/Aeolian mode, which is also determined by the key (e.g., D major). As mentioned previously, our method allows users to choose their preferred key or select a random key, where the key containing the highest alignment with the melody (termed as \textit{key confidence} in this paper) is chosen among a couple of randomly selected key signatures. A pitch range is constructed based on the selected key and kept within a certain comfortable singing range for most individuals. The phrase lengths are considered to gauge the approximate average number of measures used per phrase.

The parameters described above help the pitch generation process. This process is paired with pitch insertion, and the two processes are in a feedback loop until all pitches have been inserted into the score. The pitches would first be randomly generated but automatically adapted to music theory; then, they are inserted into the corresponding lyrics before multiple notes are adjusted by phrase to ensure a smoother connection between the notes and create a flowing melody. The degree of randomness would provide greater phrase variety even after pitch adjustment, thus preventing excess repetition that would facilitate less enjoyment. 

\subsubsection{XML Score Conversion}\label{subsec:convert to humanity}

When the score construction is complete along with the pitch construction, the score is converted to MusicXML \cite{musicxml} files for readability and easier usage in composition software and general audio/music notation software. Additionally, converting the completed score into MusicXML provides easier digital sheet music exchanges and collaborations. In the system, the MusicXML file is utilized to help display the sheet music upon the generation of non-lyrical music or a lyrical song. 

\setlength{\parskip}{0em}    
\subsection{Image-to-Lyrics Generation}

\begin{figure}[t]
    \centering
    \includegraphics[width=0.95\columnwidth]{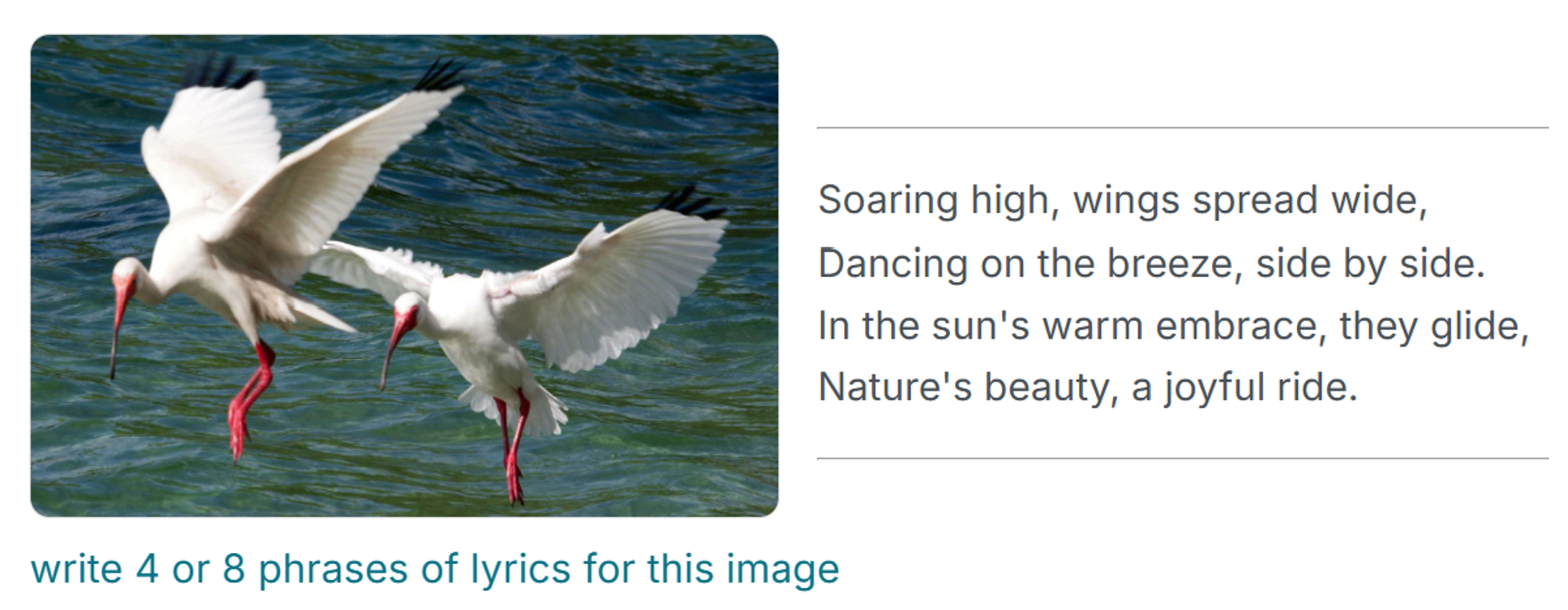}
    \caption{Lyrics generation from the image.}
    \label{fig:lyric_to_img_LLM}
\end{figure}

Different images are fed into LLMs to automatically generate lyrical rhythmic information of multiple phrases. In addition to the input images, preference prompts are added to customize the lengths generated by the LLMs. The generated lyrical rhythmic information is then employed to construct the rhythmic score. Figure \ref{fig:lyric_to_img_LLM} presents examples of prompts designed to generate song lyrics from a given image using the LLM, along with the resulting lyrics.

\subsection{GenAIM System Architecture}

\begin{figure}[t!]
    \centering
    \includegraphics[width=0.95\columnwidth]{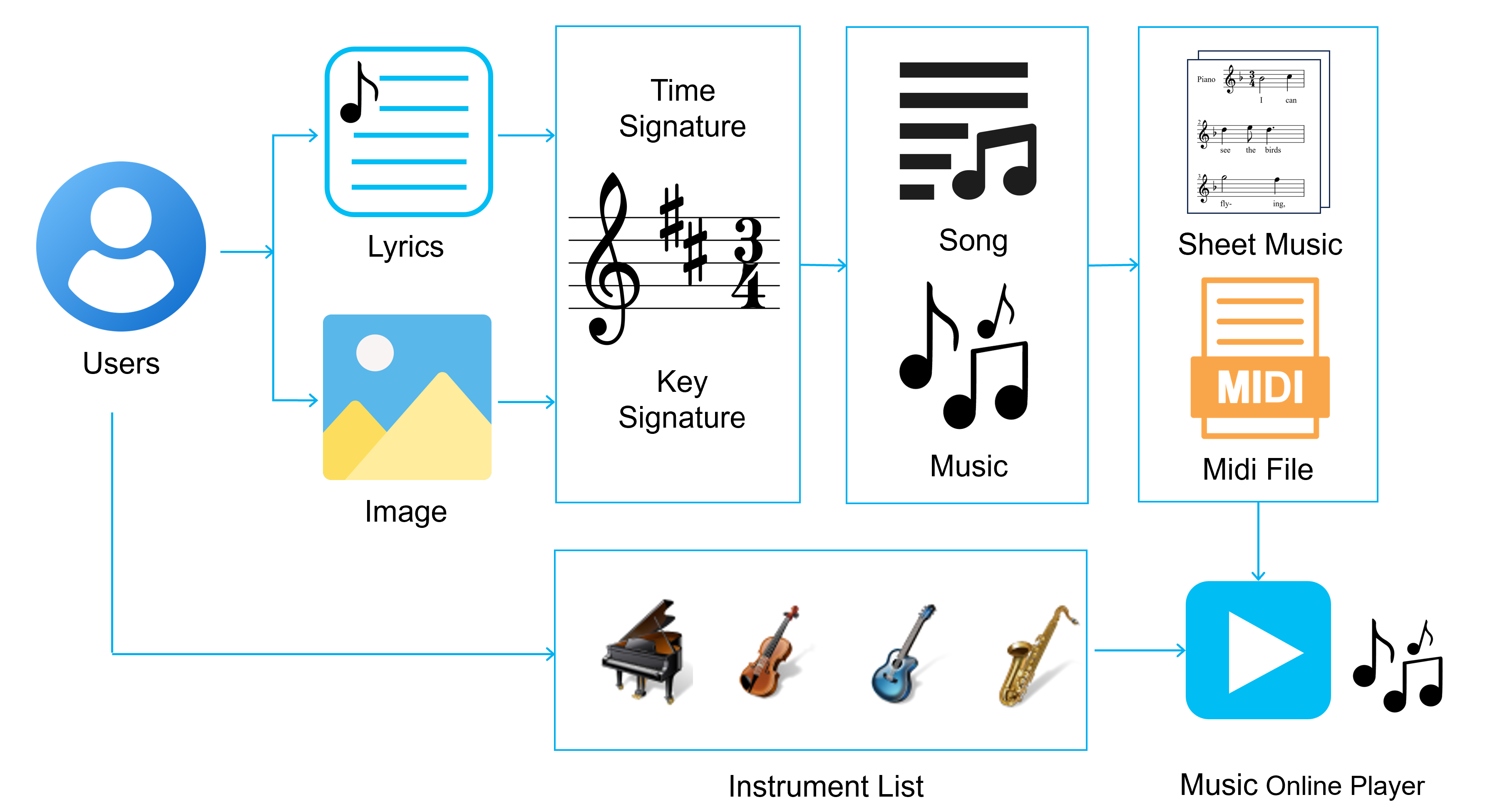}
    \caption{The user level of GenAIM.}
    \label{fig:framework_ai_music_user}
\end{figure}

\begin{figure*}
 \centerline{
 \includegraphics[width=0.80\textwidth]{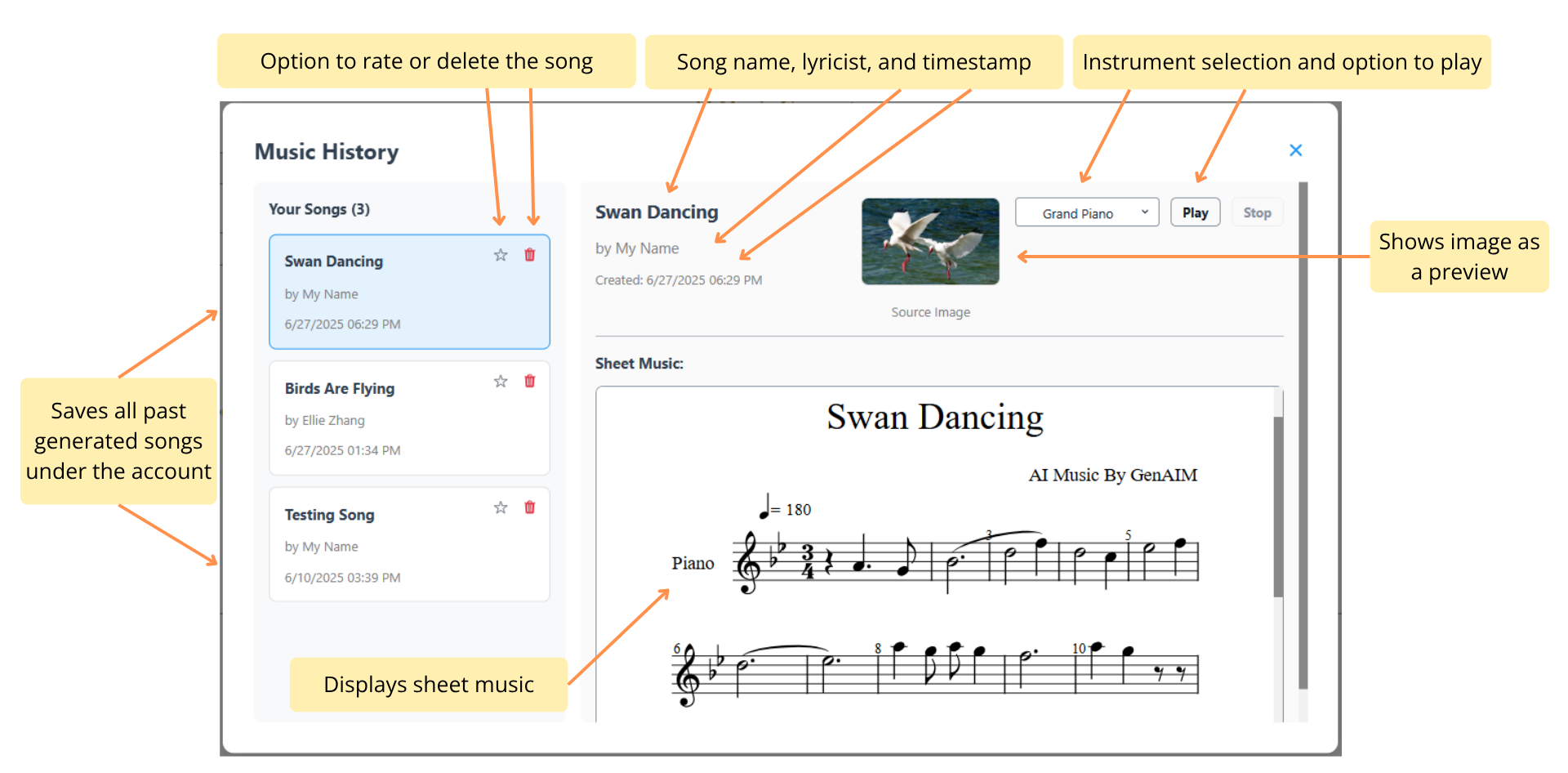}}
 \caption{Display of the music history feature in GenAIM.}
 \label{fig:genAIM_music history}
\end{figure*}

This MusicAIR framework is built on the Amazon Web Services (AWS) platform. It utilizes several services including networking and content delivery, application integration, database, storage, artificial intelligence, security, identity, and compliance. The core music generation module directly creates music from the input lyrics, producing both sheet music and audio files. The input image is processed by the LLMs of the AI service, which generate the lyrics \cite{IMAiGen2025}. These generated lyrics are then passed to the music core for music composition. The front end web application loads the music files from the AWS cloud, renders to a music sheet, and plays the music.

The GenAIM system architecture is explained through the user level and the system level. 

\subsubsection{User Level}
As shown in Figure \ref{fig:framework_ai_music_user}, two forms of input are accepted at the user level: lyrics and image. Users can choose their preferred key signature or the "random" option that selects a moderately fitting key, while the most fitting time signature is determined by the algorithm based on the input lyrics. Additionally, users can choose to generate non-lyrical music or a lyrical song. When the result has been generated, a dropdown menu offers a variety of instruments that the user can select for playback, such as the guitar and piano. Additionally, users can access their music history to view all of the past generated songs, rate their songs, and replay and review them again if necessary. The music history is shown in detail in Figure \ref{fig:genAIM_music history}.

\subsubsection{System Level}

At the system level shown in Figure \ref{fig:MusicAIR_sys}, an image or a set of lyrics is accepted, but the features are extracted from the image using LLMs before further processing is performed. The lyrics or image features are transferred to the music core, where pitches, rhythms, phrases, etc. are determined based on the lyrics. Upon generating non-lyrical music or a song, the framework presents the results in a MIDI or sheet music format.

\vspace{-0.5em}  

\subsection{Evaluation Metrics}

Multiple metrics are used to evaluate the robustness of the system. The main metrics are key confidence, melodic smoothness, and rhythm matching. Key confidence evaluates key signature alignment with the generated melody, while melodic smoothness evaluates the degree of resemblance to human-composed melodies through the smoothness of a melody. Rhythm matching evaluates alignment between stressed beats and stressed lyrical elements and vice versa.   

\subsubsection{Key Confidence}

In this paper, the term \textit{key confidence} is defined by the extent to which a musical passage implies a particular key based on human perception. This metric is denoted as the correlation coefficient $r$:

\begin{equation}
r = \frac{\sum_{i=1}^{n} (X_i - \bar{X})(Y_i - \bar{Y})}
{\sqrt{\sum_{i=1}^{n} (X_i - \bar{X})^2 \sum_{i=1}^{n} (Y_i - \bar{Y})^2}}
\label{eq:key_confid}
\end{equation}

In Equation \ref{eq:key_confid}, $X_i$ and $\bar{X}$ represent the input vector and its corresponding average, respectively; $Y_i$ and $\bar{Y}$ represent the key-profile values of the designated key and its corresponding average, respectively; and $n$ is the number of pairs. The Krumhansl-Schmuckler key-finding algorithm \cite{KS1982original} is employed to calculate the correlation coefficient \cite{temperley1999KS}. The algorithm utilizes a set of key profiles derived from human perception of the importance of each pitch within a key for all 24 scales. These key profiles are used to compare with the input music to calculate the correlation, and the key with the highest correlation is the most fitting key for the music. 

\subsubsection{Melodic Smoothness}
Melodic smoothness metrics include average interval, step ratio, and direction change rate. 

The \textit{average interval} is defined as $s$ in Equation \ref{eq:rel_smooth}, where $|\Delta x_{i}|$ represents each absolute interval between two successive pitches and $N$ represents the total number of intervals. A lower average interval represents a smoother melody, reflecting smaller pitch deviations between notes. 

\begin{equation}
s = \frac{\sum_{i=1}^{N} |\Delta x_{i}|}{N}
\label{eq:rel_smooth}
\end{equation}

The \textit{step ratio} is a complementary metric for evaluating melodic smoothness. In music theory, step are characterized by intervals of one or two semitones (i.e., half steps) between successive pitches, while leaps exceed this range \cite{openmusictheory}. A higher step ratio consisting of more steps than leaps indicates more stepwise motion. However, even melodies composed entirely of steps can potentially lack natural, human-like variation, since smaller leaps still occur commonly in human-composed melodies. As a result, it is critical to include both metrics in the evaluation of melodic smoothness. The step ratio is defined in Equation \ref{eq:step_ratio}:

\begin{equation}
\textit{Step-Ratio} = \frac{\textit{steps}}{\textit{steps} + \textit{leaps}}
\label{eq:step_ratio}
\end{equation}

The \textit{direction change rate} represents contour smoothness and counts how often the direction of the melody changes. Fewer direction changes correspond to a smoother and more flowing melody. The rate is normalized by the sequence length to get a relative value.

\begin{figure}[b]
 \centerline{
 \includegraphics[width=0.7\columnwidth]{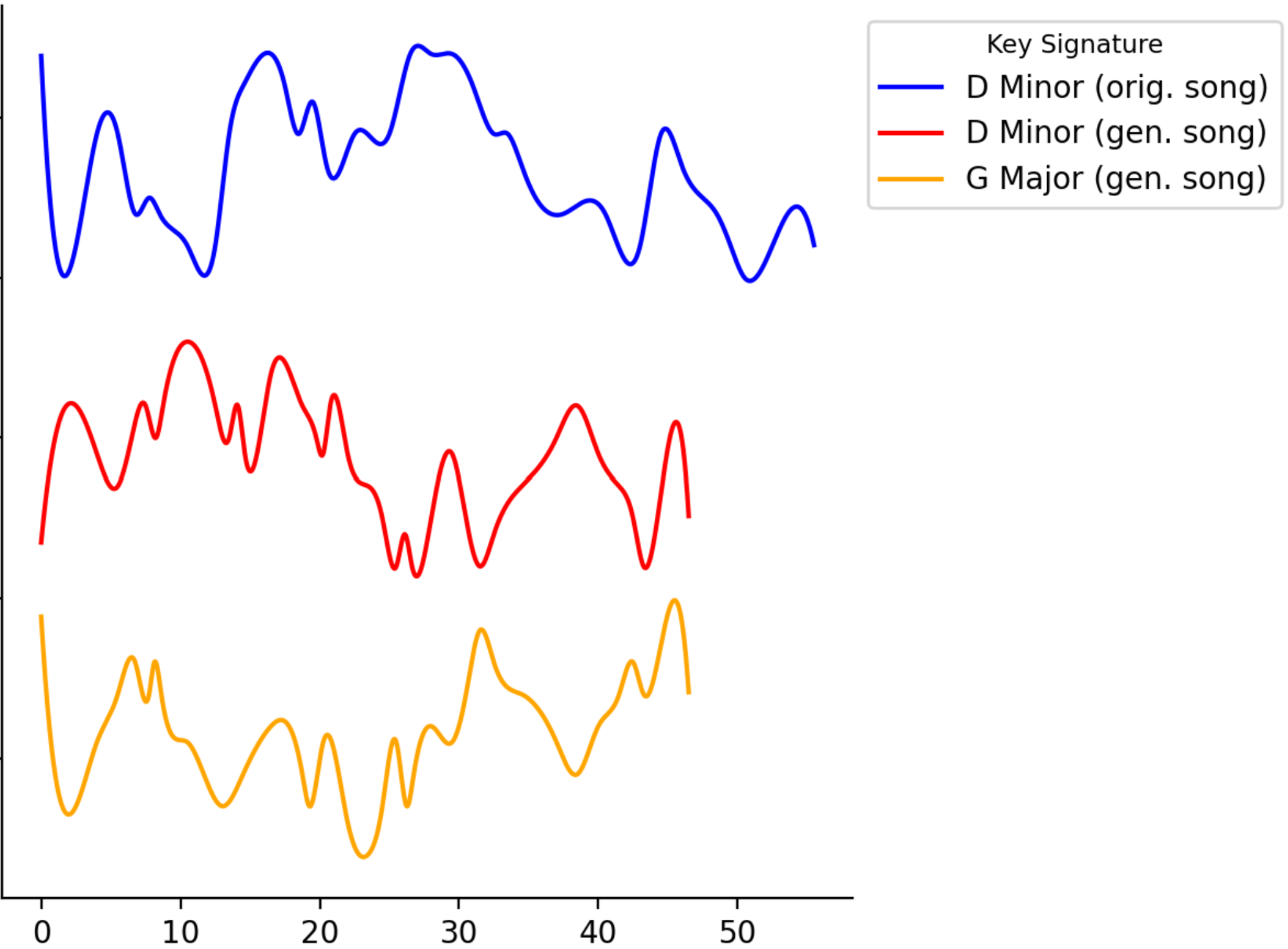}}
 \caption{Melody motion comparisons.}
 \label{fig:genAIM_melody_motion}
\end{figure}

\section{Experimental Results \& Evaluation}\label{sec: experiments}

\begin{figure*}
    \centering
    \begin{subfigure}[t]{0.5\textwidth}
        \centering
        \includegraphics[width=0.95\columnwidth]{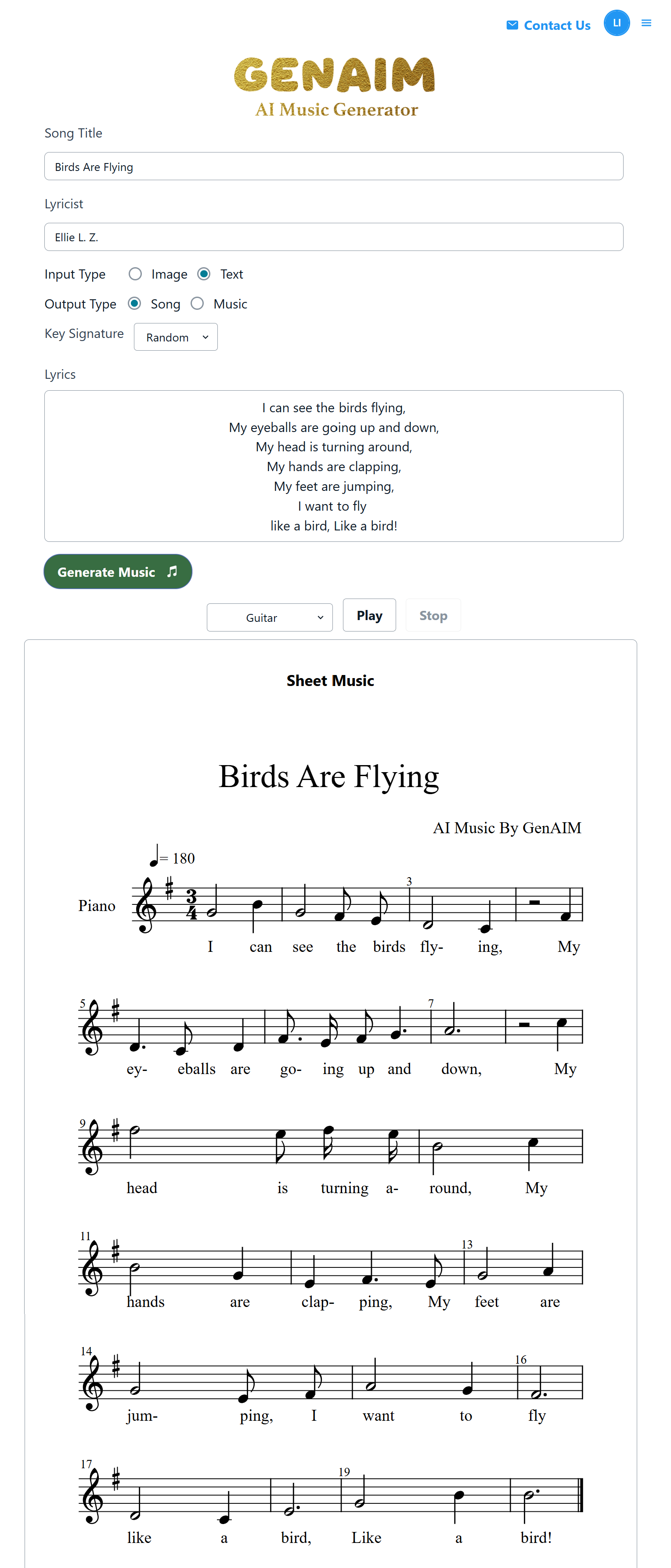}
        \caption{Lyric-to-music generation.}
        \label{fig:genAIM_img2music}
    \end{subfigure}%
    ~
    \centering
    \begin{subfigure}[t]{0.487\textwidth}
        \centering
        \includegraphics[width=0.98\columnwidth]{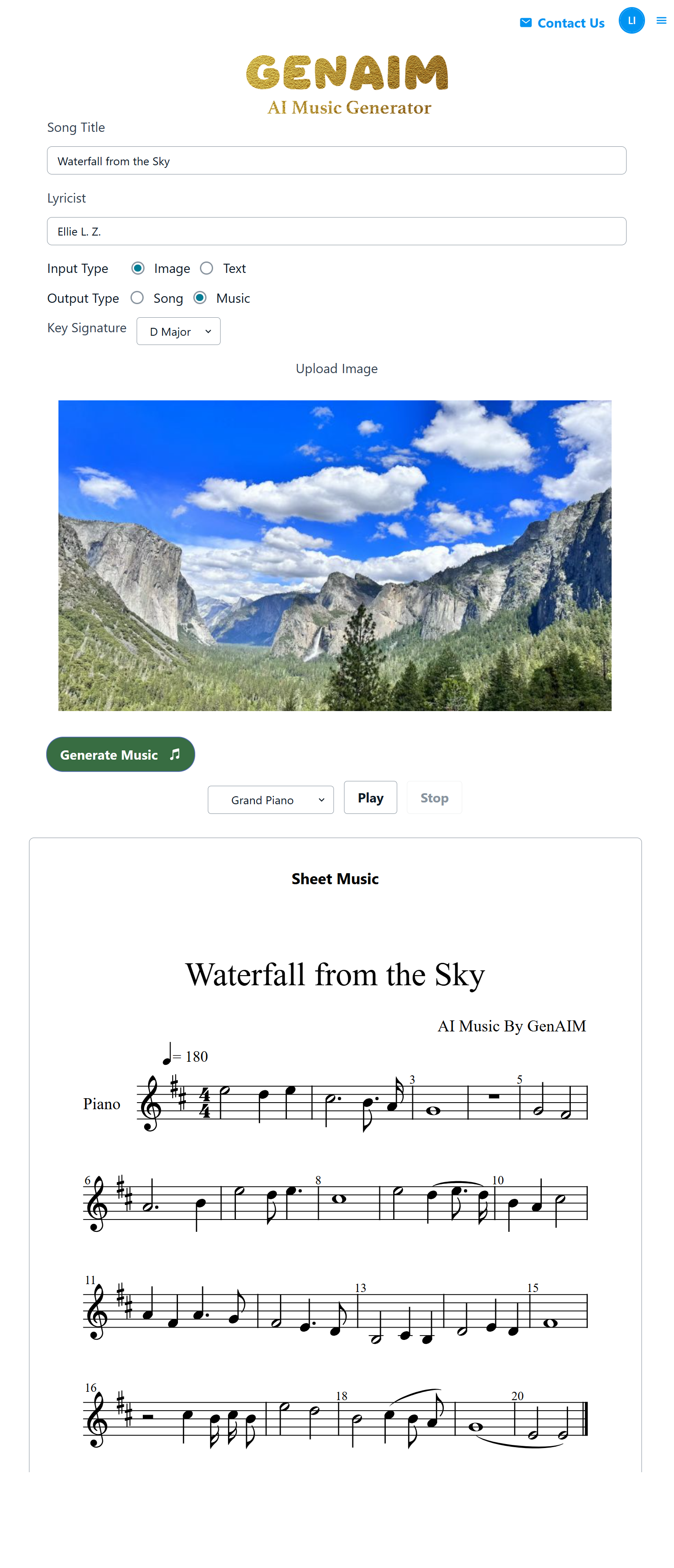}
        \caption{Image-to-music generation.}
        \label{fig:genAIM_lyrics2music}
    \end{subfigure}%
    \caption{Music generation from lyrics and the image.}
    \label{fig:lyric_image_gen}
\end{figure*}

We performed AI music generation with the web tool GenAIM using our proposed framework, MusicAIR, and conducted exploratory data analysis using Music21 \cite{Music21}, a Python-based development toolkit created by the Massachusetts Institute of Technology (MIT), to discover potential similarities and differences between an AI-generated and an original composition. The toolkit has been established as robust for theory-based evaluations and is suitable for the field of algorithmic composition \cite{CuthbertA10}. Human listening evaluations are not included in this paper, as the core focus is on theory-based metrics and the background variability of listeners may create conclusions that are difficult to consolidate. Additionally, we seek to discover correlations and trends within each composition. The web tool GenAIM generates music from text and image forms, as seen in Figures \ref{fig:genAIM_music history} and \ref{fig:lyric_image_gen}. The algorithm-driven music core has no specific requirements for GPU/CPU models, the amount of memory, and the operating system. The GenAIM system is deployed on AWS, and only Python 3.0+ is used for implementation.
 
Two experiments are performed: 1) 24 original songs from children's piano books are used for comparison with multiple AI-generated scores for each song; and 2) the evaluation metrics evaluate 517 generated music from the GenAIM database by anonymous users. Sections A to C cover the first experiment, and Section D covers the second experiment.

\begin{figure*}
    \centering
    
    \begin{subfigure}[t]{0.45\textwidth}
        \centering
       \includegraphics[width=0.98\columnwidth]{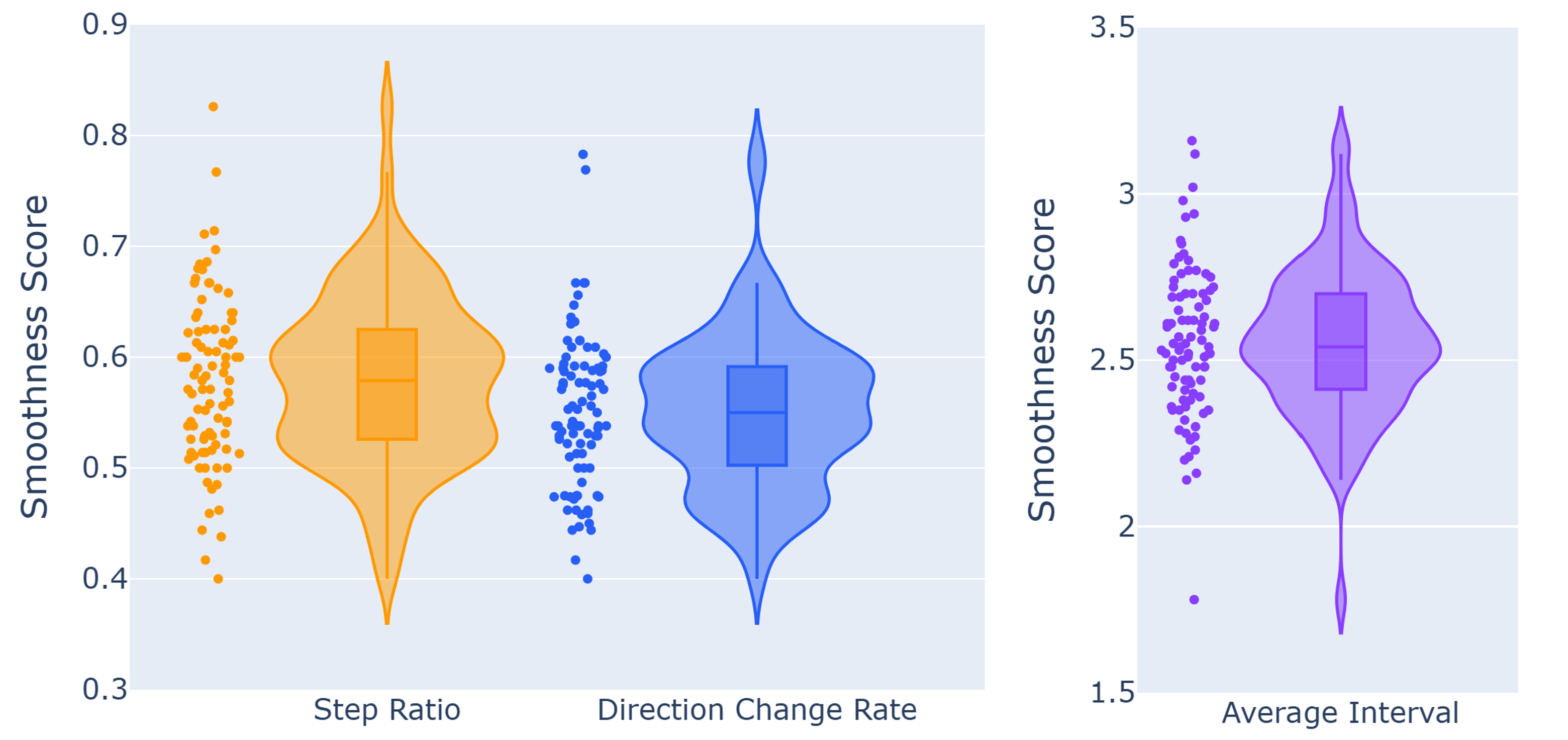}
        \caption{Melodic smoothness of generated songs.}
        \label{fig:genAIM_smoothness_gen}
    \end{subfigure}%
    ~
    \begin{subfigure}[t]{0.5\textwidth}
        \centering
       \includegraphics[width=0.88\columnwidth]{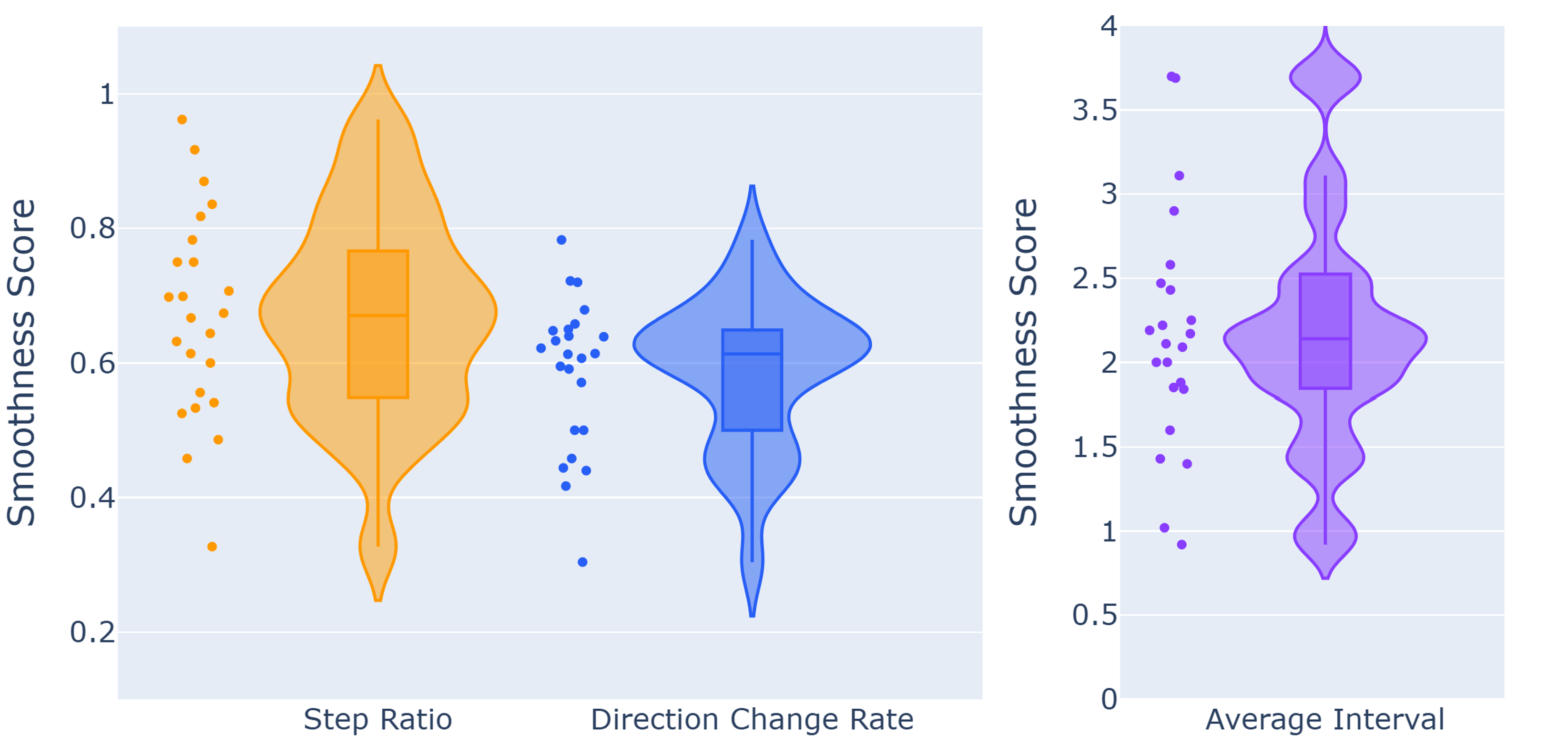}
        \caption{Melodic smoothness of original songs.}
        \label{fig:genAIM_smoothness_orig}
    \end{subfigure}
    
    \caption{Comparison of Melodic smoothness.}
    \label{fig:comp_smoothness}
\end{figure*}

\begin{figure*}[ht]
    \centering
    
    \begin{subfigure}{0.32\textwidth}
        \centering
        \includegraphics[height=1.5in]{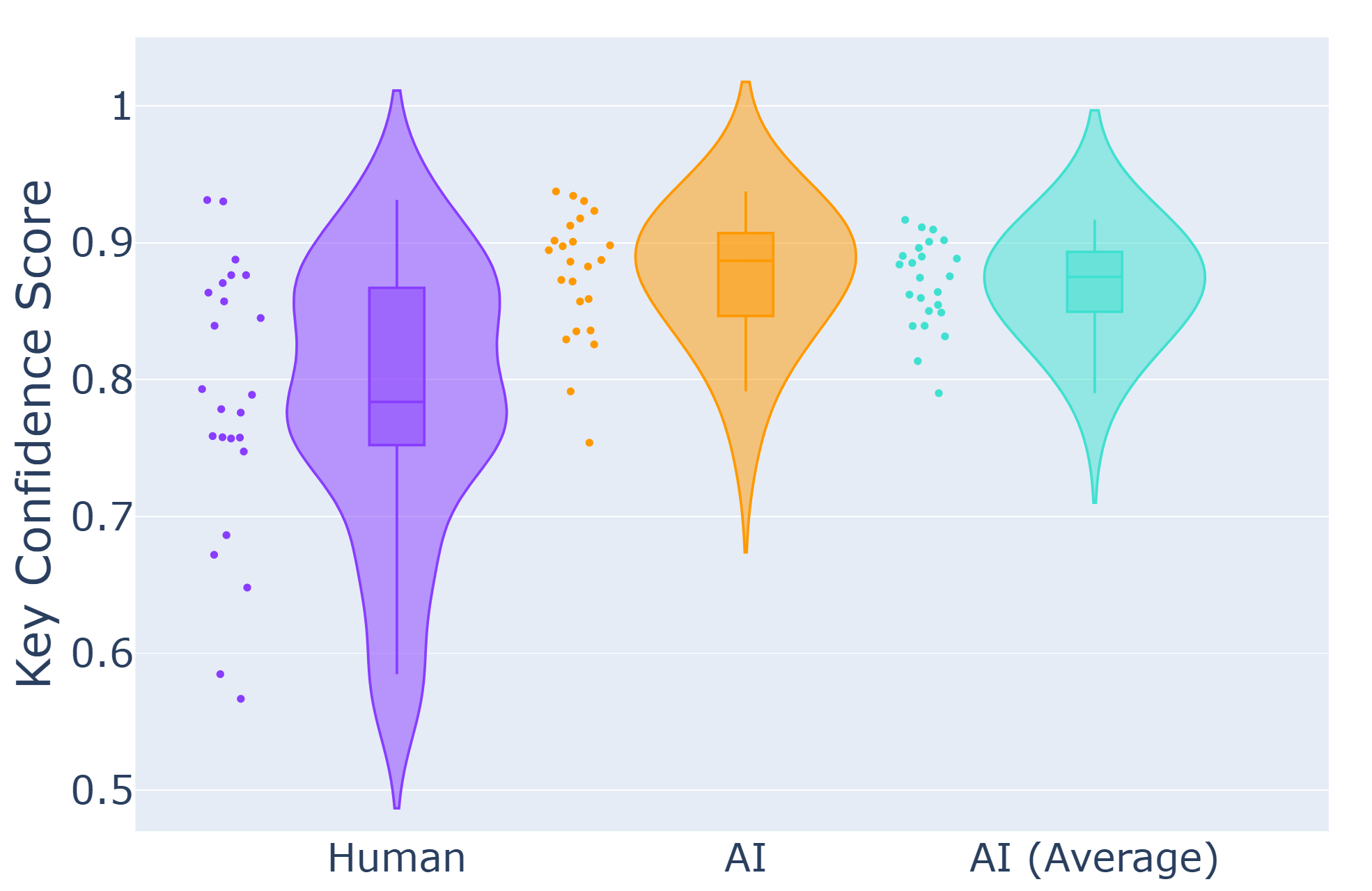}
        \caption{Key confidence comparisons.}
        \label{fig:genAIM_violin_boxplots}
    \end{subfigure}%
    ~
    \centering
    \begin{subfigure}{0.34\textwidth}
        \centering
        \includegraphics[height=1.5in]{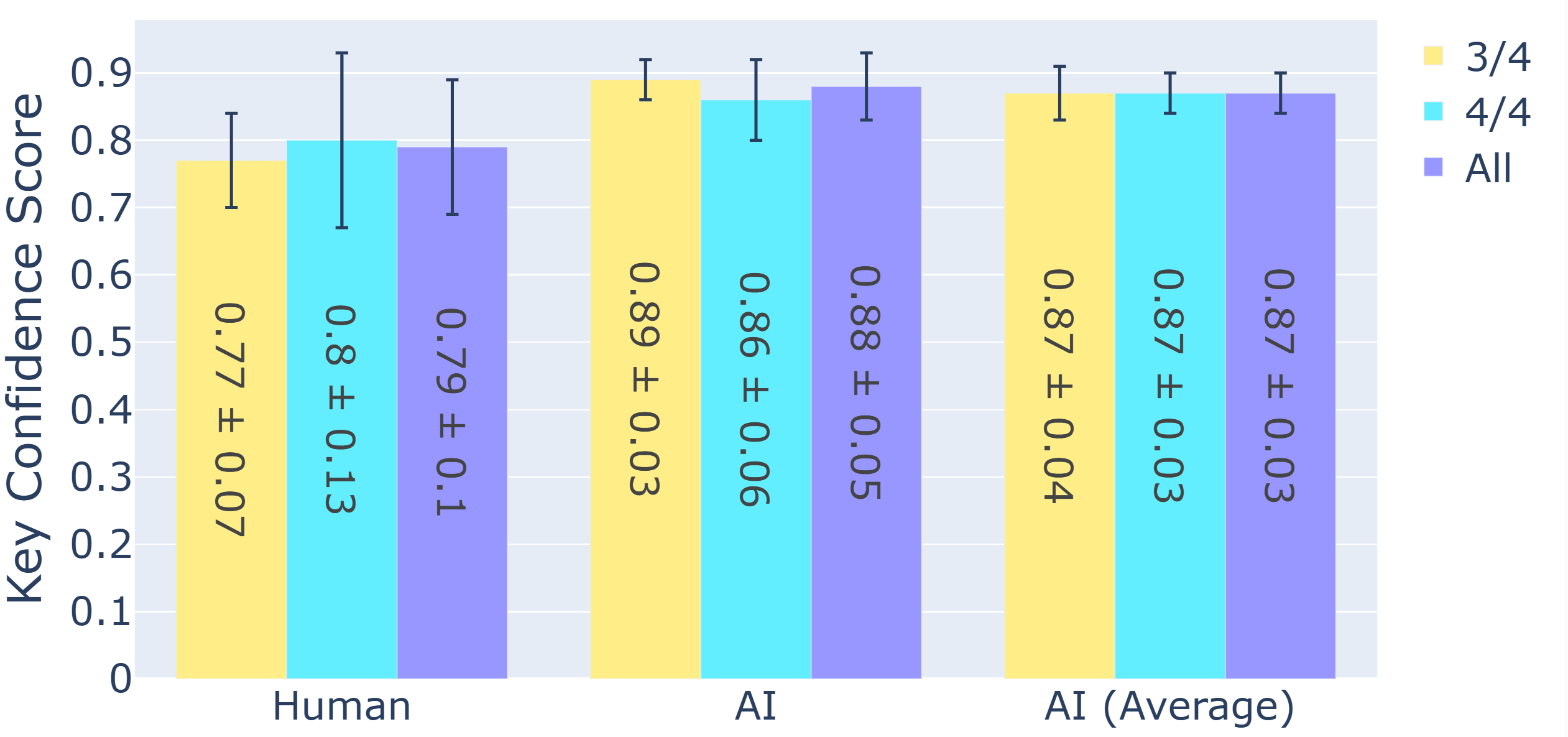}
        \caption{Average Key confidence comparisons.}
        \label{fig:genAIM_key_confidence}
    \end{subfigure}%

    \caption{Evaluations of the paired original and AI-generated songs.}
    \label{fig:comp_pairs}
\end{figure*}

For the first experiment, all AI musical scores are generated by the music core using the same lyrics from the songs. For each song, 3 to 5 songs are generated in the same time signature with different keys from a selection of 17 key signatures, resulting in the collection of 97 AI-generated songs. It is guaranteed that for each set of lyrics, there will be at least one generated song that contains the same key signature as the original song. Figure \ref{fig:Two compositions} displays the two compositions for comparison. Figure \ref{fig:EllieLZ_score_strip} displays an original song written by a young child \cite{birdsareflying}. Figure \ref{fig:AI_score_strip} showcases one version of the AI-generated musical score that utilizes the same lyrics. The first experiment is expanded upon further in the upcoming sections.

\subsection{Melodic Motion and Smoothness}

Figure \ref{fig:genAIM_melody_motion} shows a motion plot of three example AI-generated songs in different key signatures. On the x-axis, the numbers are relative to the number of notes generated throughout the piece. The figure shows that each turning point in the generated songs closely aligns with those of the original song, revealing that the rhythmic patterns between the original and generated songs are highly similar. Although there is more deviation from the original song in the pitch trends, this is reasonable because pitches have more flexibility in determining melodic flow as long as the melody is not too jarring or sudden. Additionally, the general motions of all five songs are visually smooth, and there are no jagged portions, displaying a closer resemblance to human-composed music.

Figure \ref{fig:genAIM_smoothness_gen} presents violin-box plots of smoothness metrics for generated songs. The average interval represents the mean absolute pitch distance between successive notes, with lower values corresponding to more stepwise motion. The step ratio measures the proportion of stepwise to leaping motion and is generally associated with smoother melodies, and such melodies often have values above 0.6. Lastly, the direction change rate evaluates the frequency with which a song changes direction melodically, where high values would indicate too much variability. In the figure, the median average interval of 2.58 is consistent with typical stepwise motion due to its closeness to the interval range for a step. The median step ratio is 0.574, which approaches the threshold 0.6. The median direction change rate of 0.567 indicates a balance in the frequency of changing direction. Overall, these metrics reveal that the generated songs display high melodic smoothness and show that the algorithm executes a smooth musical pitch construction.

Figure \ref{fig:genAIM_smoothness_orig} shows violin-box plots of smoothness metrics for the original songs. Compared to the generated songs, the original songs exhibit higher median step ratios and direction change intervals, but lower average intervals. The higher step ratio suggests more steps than leaps, while the lower average interval indicates less distance between successive notes. However, the higher direction change rate reflects a higher chance of jarring changes, and the overall uneven and stretched distribution for the original songs indicates high variability. After comparisons, the generated songs still maintain consistent high melodic smoothness.

\vspace{0em}  

\subsection{Key Confidence}

Figure \ref{fig:genAIM_violin_boxplots} displays violin-box plots of key confidence scores that compare scores between three composer types: human, AI, and averaged AI. The key confidence score uses the key analysis algorithms demonstrated in Equation \ref{eq:key_confid} to determine the tonal center. The AI composer generates a song retaining the same time signature and key signature that the human composer used for every shared song, but the averaged AI composer generates 3-5 song versions per set of lyrics without needing to adhere to the human composer's selected parameters. From the figure, the distribution for the human composer is significantly broader and longer than the distribution for both AI composers. The AI composers have a more concentrated \textit{key confidence} that is especially in the 0.8-0.9 range---0.887 and 0.876 for the median of the AI and AI averaged composers, respectively, indicating that an AI composer is more likely to generate songs with a more fitting key signature than a human composer.

\begin{figure}[h]
    \centering
    \includegraphics[width=0.50\columnwidth]{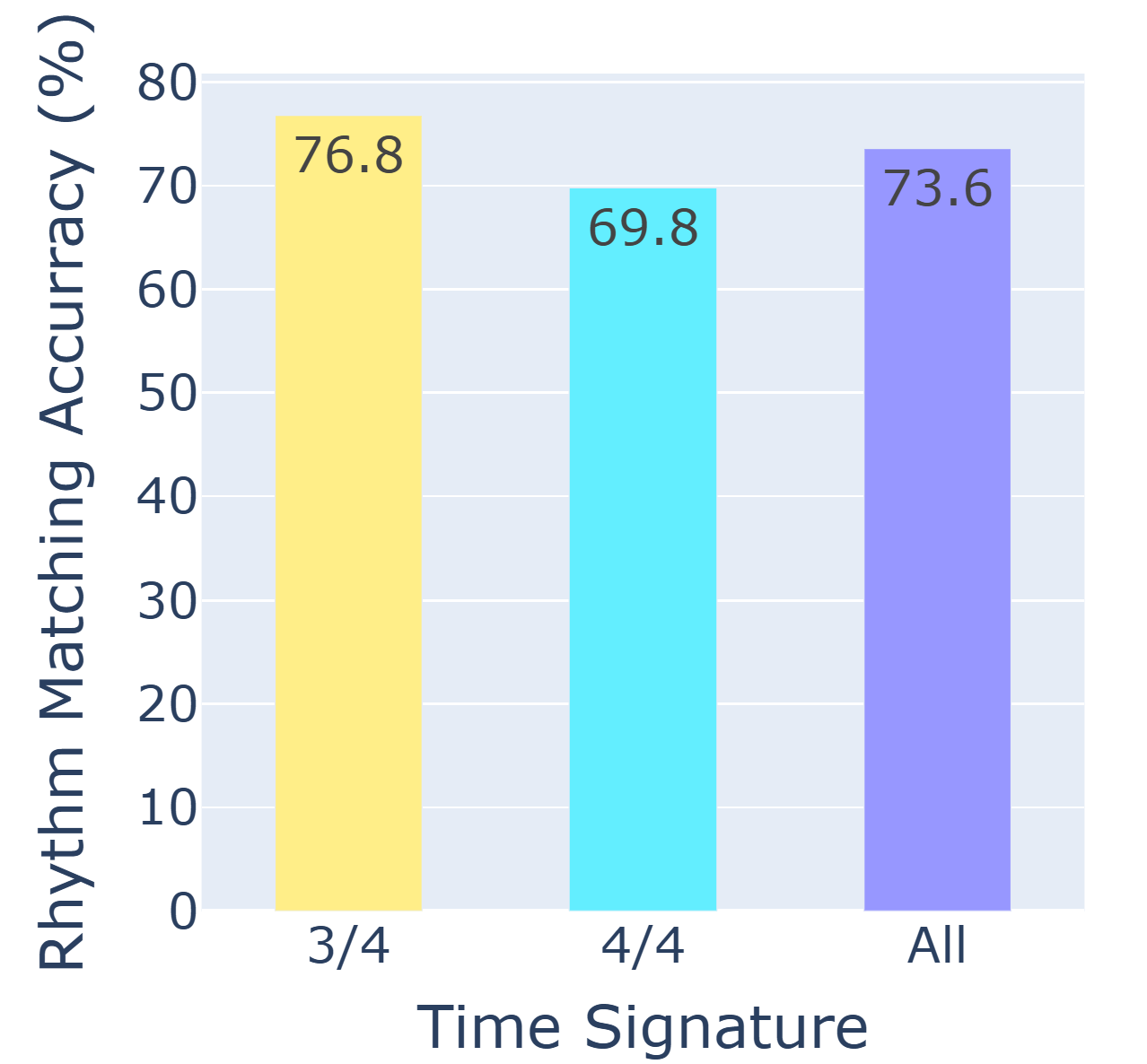}
    \caption{Rhythm matching comparisons.}
    \label{fig:genAIM_rhythm_matching}
\end{figure}

\begin{figure*}
    \centering
    \begin{subfigure}[t]{0.33\textwidth}
        \centering
        \includegraphics[height=1.7in]{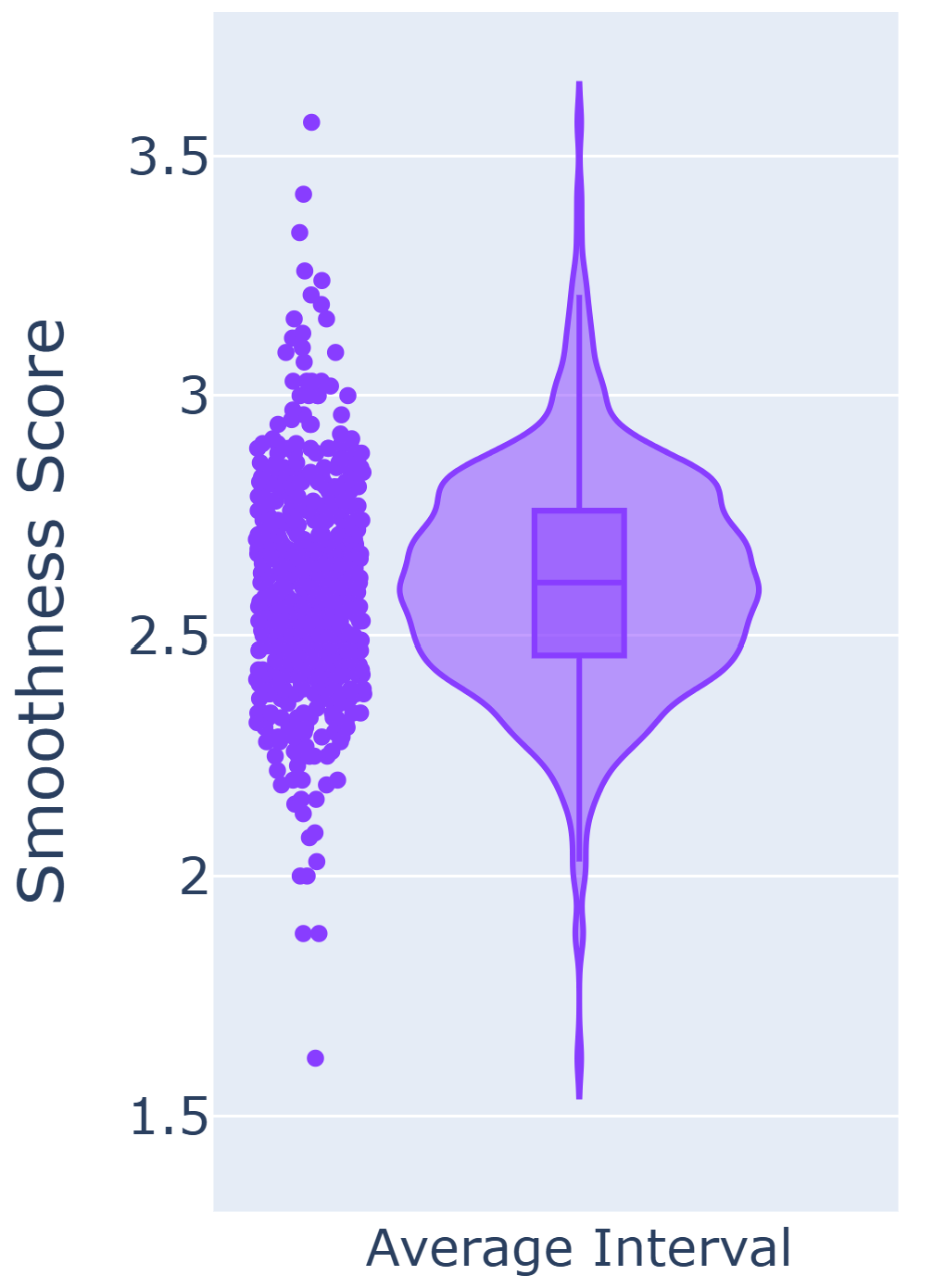}
        \caption{Average Interval}
        \label{fig:genAIM_517_smoothness_gen}
    \end{subfigure}%
    ~
    \centering
    \begin{subfigure}[t]{0.35\textwidth}
        \centering
        \includegraphics[height=1.7in]{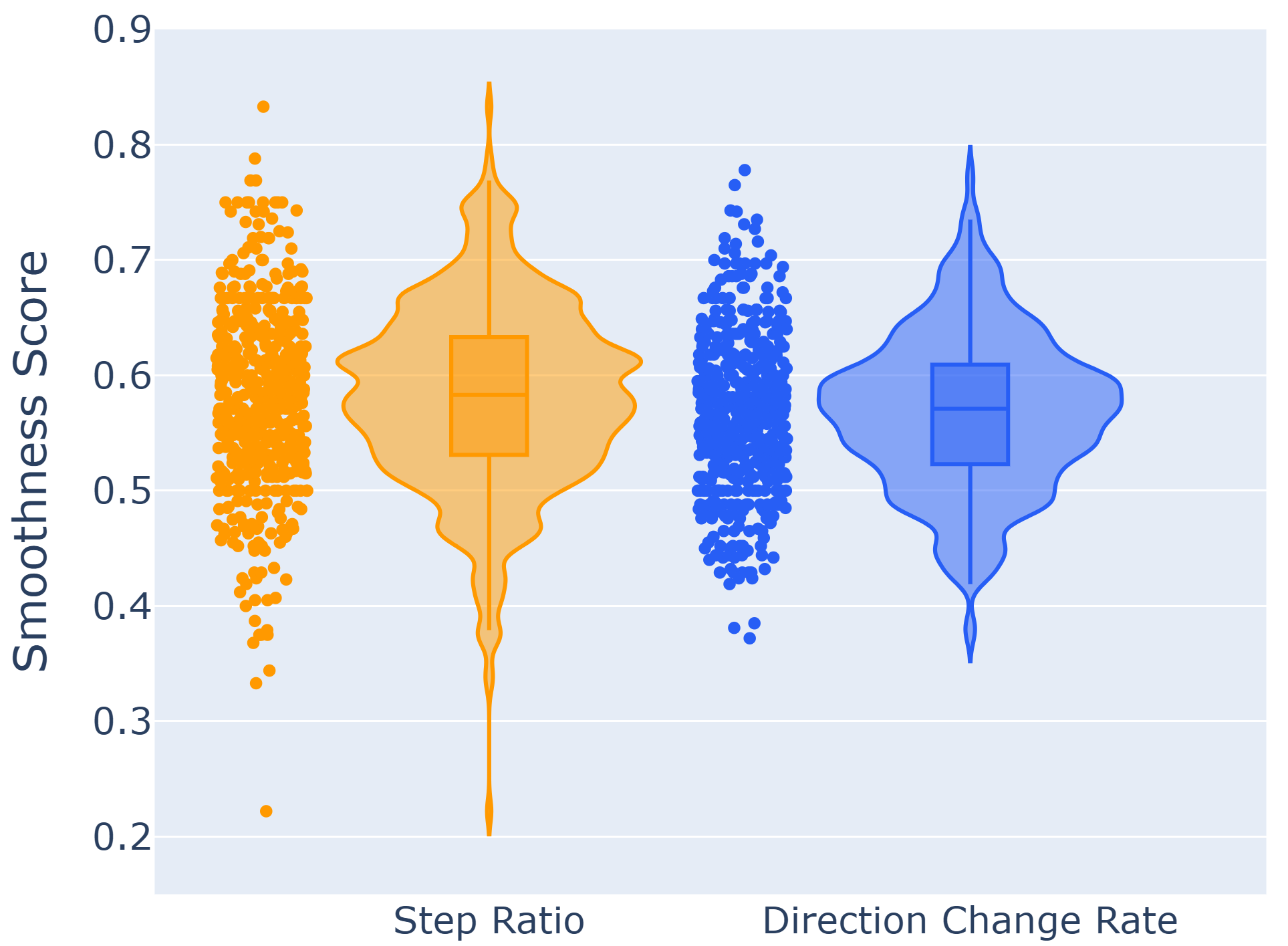}
        \caption{Step Ratio and Direction Change Rate}
        \label{fig:genAIM_517_smoothness_gen2}
    \end{subfigure}%
    ~
    \begin{subfigure}[t]{0.37\textwidth}
        \centering
        \includegraphics[height=1.7in]{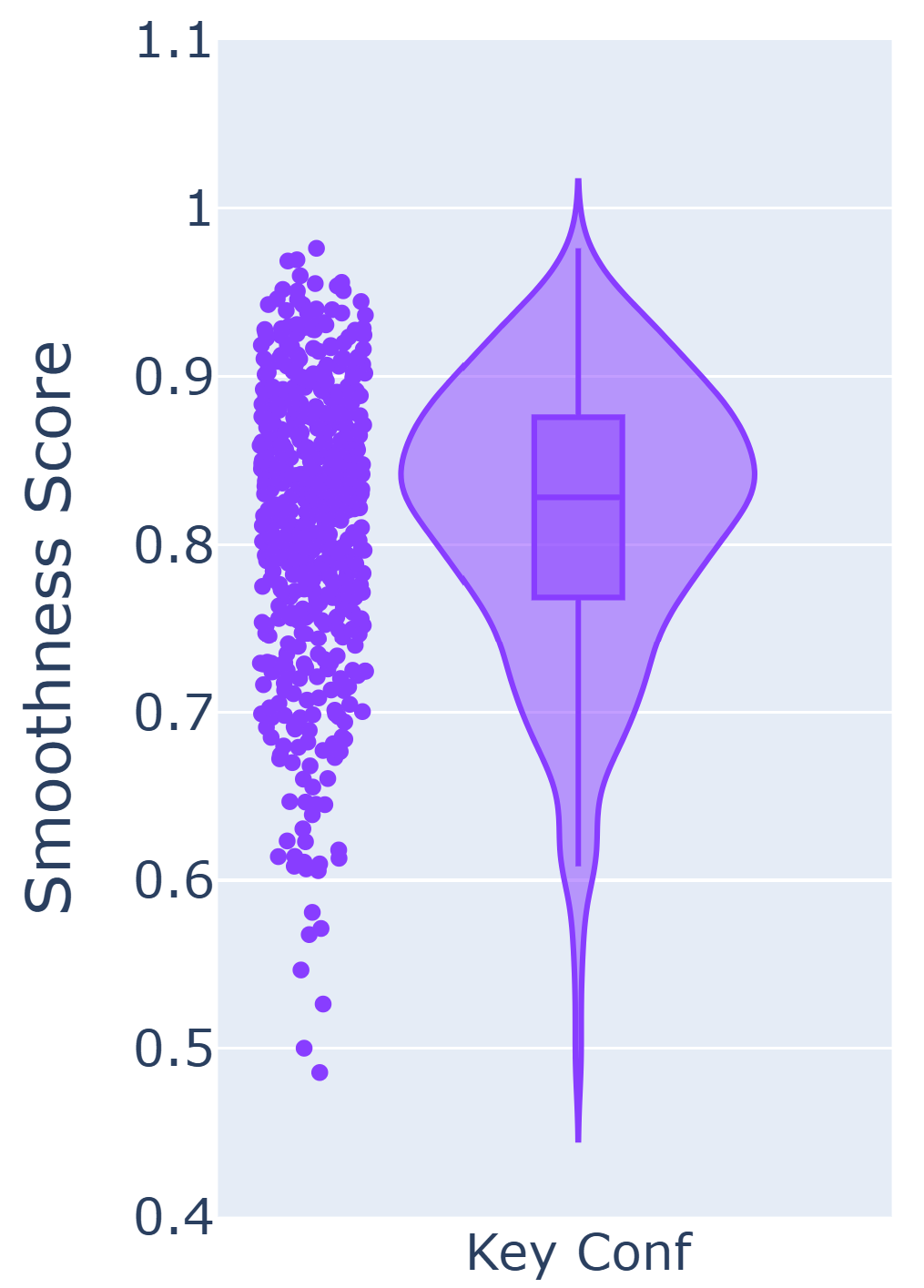}
        \caption{Key Confidence}
        \label{fig:genAIM_517_key_confidence}
    \end{subfigure}

    \caption{Evaluations of AI-generated 517 songs from both lyrics and images by anonymous users.}
    \label{fig:more_demos}
\end{figure*}

Figure \ref{fig:genAIM_key_confidence} displays similar information, except that each composer is divided by time signatures. Key confidence scores are averaged based on the time signature (3/4, 4/4, or all). Between the AI composers, the regular AI composer still retains the same key signature that the human composer selected, but the averaged AI composer does not need to. In this figure, all the human-composed songs have average key confidence values around 0.80, while the AI composers all have a minimum of 8\% higher key confidence value than the human composer in their respective time signature, reinforcing our method's reliability in producing fitting melodies. 

\subsection{Rhythm Matching}

Figure \ref{fig:genAIM_rhythm_matching} displays rhythm matching, which is critical for ensuring that rhythms give more priority through durations and beat locations to keywords. For evaluating the average rhythm matching accuracy of generated songs when compared to original songs, the score was 76.8\% for 3/4 time signatures and 73.6\% for all songs regardless of time signatures, indicating a relatively high matching accuracy. Although the rhythm matching score was 69.8\% for 4/4 time signatures, deviation from the original song is expected, as song composition is not definite and contains a significant degree of flexibility --- it is only recommended that keywords and strong beats are associated with each other based on research from \cite{Liao2022Multimodal}\cite{CLiao2024}.

\vspace{0em}  

\subsection{Comparison Metrics with User-Generated Songs}

In this alternate experiment with 517 generated music from both lyrics and images by anonymous users, the same evaluation metrics are used. However, because the music originates from anonymous contributors and the lyrics or images have not previously been set to music, there are no original songs or music available for comparison. Figures \ref{fig:genAIM_517_smoothness_gen}, \ref{fig:genAIM_517_smoothness_gen2}, and \ref{fig:genAIM_517_key_confidence} display the corresponding violin plots for the average interval, the step ratio and direction change rate, and the key confidence, respectively. For median values, the average interval of 2.56 is still consistent with stepwise motion; the step ratio of 0.583 is close to the threshold 0.6 as well; the direction change rate of 0.571 also suggests that there is a relatively equal frequency of changing to different directions; and the key confidence value of 0.828 is similarly high, demonstrating high key signature alignment with the generated melodies. As a result, this indicates that GenAIM's performance is consistent across all generated songs and aligns with theory-based evaluation benchmarks. 

\section{Conclusions \& Future Work} 

\setlength{\parskip}{0em}    

We introduced MusicAIR, a state-of-the-art multimodal music generation framework powered by an algorithm-driven lyric-to-music generation core. The method supported by our framework enables music generation by integrating both lyrical and non-lyrical inputs, including text and images. Furthermore, the lyric-to-music generation process does not rely on prior training data. Comprehensive musical analysis performed through music21, a confirmed robust method of analysis and evaluation through music theory, compared the results of 97 AI-generated songs to the original human-composed songs, as well as evaluated 517 anonymously generated music from the GenAIM database. Our experimental findings demonstrate that the overall averaged key confidence values for the generated songs matching one original song reach 0.85 compared to 0.79 for the original, human-composed songs. The average rhythmic matching is 73.6\% for all generated songs when compared with original songs. Furthermore, the anonymously generated songs demonstrate a 0.828 key confidence. Therefore, using MusicAIR, GenAIM is capable of generating highly fitting melodies rhythmically and melodically. Although human listening evaluations were not performed, we believe that evaluating based on music theory standards, an objective and reasonable nature, is more fitting for the research in this paper in particular than listening evaluations, as the main objective of our paper is to determine if GenAIM aligns with music theory rules. Overall, our novel method can inspire music generation techniques from a refreshing perspective. As our framework evolves, we believe our method will benefit both amateur and professional musicians as a copilot and an educational tutor and help accelerate the composition workflow and learning process by aspiring musicians.

Since this paper is solely focused on the generation of main melodies, no further musical complexity is added. For future work, the pitch construction can still be improved upon by adding more layers of complexity, such as including chord progressions and a variety of cadences. Furthermore, more experiments with different songwriting genres can be considered to present more diverse AI-generated works, which may impact both the rhythmic structure and pitch construction.

\section*{Limitations}
One limitation of using LLMs to extract image information is that the process may lack contextual depth and creativity. Furthermore, our algorithm for generating music from lyrics currently does not have multiple variations for different moods and genres. These limitations will be improved upon and optimized in our future work. 

\section{Ethical Implications}

Lyrics-to-music generation with deep learning poses ethical challenges. The proposed method does \textit{not} use deep learning in the lyric-to-music process, but LLMs are utilized to analyze the image for music inspiration. LLMs help shape rhythmic structure but do not directly generate music, preventing copyright concerns and ensuring the AI-generated songs are original. All images, text, lyrics, and music samples used are either created by the authors or generated through their GenAIM web tool, eliminating copyright issues.

\bibliographystyle{IEEEtran}
\bibliography{musicAIR}

\end{document}